# Single-shot wide-field topography measurement using spectrally multiplexed reflection intensity holography via space-domain Kramers–Kronig relations

CHUNGHA LEE,[1,2] YOONSEOK BAEK,[1,2] HERVE HUGONNET[1,2], AND YONGKEUN PARK[1,2,3*]

[1]Department of Physics, Korea Advanced Institute of Science and Technology (KAIST), Daejeon 34141, Republic of Korea.
[2]KAIST Institute for Health Science and Technology, KAIST, Daejeon 34141, Republic of Korea.
[3]Tomocube Inc., Daejeon 34109, Republic of Korea.
*Corresponding author: yk.park@kaist.ac.kr

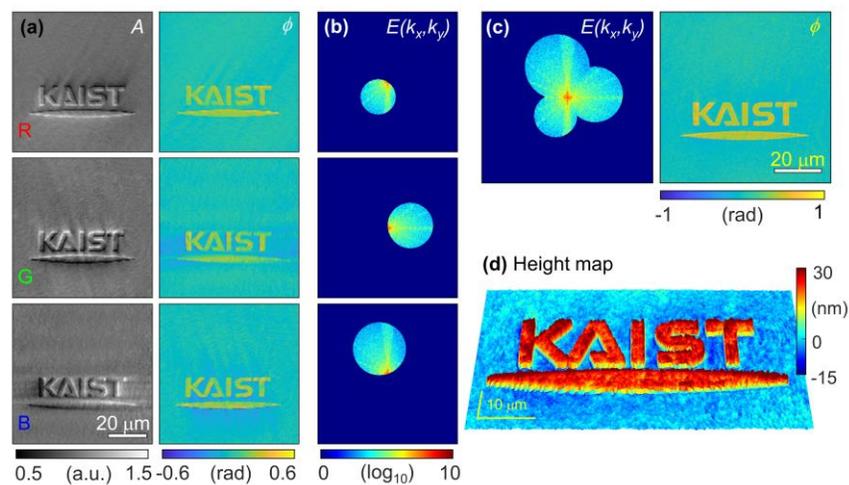

**Abstract**

**Surface topology measurements of micro- or nanostructures is essential for both scientific and industrial applications. However, high-throughput measurements remain challenging in surface metrology. We present a single-shot full-field surface topography measurement using Kramers–Kronig holographic imaging and spectral multiplexing. Three different intensity images at different incident angles were simultaneously measured with three different colors, from which a quantitative phase image was retrieved using spatial Kramers–Kronig relations. A high-resolution topographic image of the sample was then reconstructed using synthetic aperture holography. Various patterned structures at the nanometer scale were measured and cross-validated using atomic force microscopy.**

Characterization of surface topography provides essential information about the functionalities of a material with a micro- or nanostructured surface [1]. Surface characterization has been widely achieved by stylus profilometry [2], scanning electron microscopy [3], and atomic force microscopy (AFM) [4], which enables high-resolution surface topography measurements. However, real-time inspection of a surface using these techniques is challenging because of their slow scanning speed and potential invasiveness. While AFM is less invasive depending on the scanning mode, stylus profilometry uses the stylus tip in physical contact with the surface during measurement [5]. Meanwhile, scanning electron microscopy for height sensing typically requires sample cutting.

Optical surface metrology techniques have gained attention because of their non-contact and fast measurement capabilities. In particular, quantitative phase imaging (QPI) provides full-field measurements of micro- or nanostructures. In QPI, phase information is measured using light interference and used to reconstruct the surface topography. QPI allows non-destructive and high-throughput characterization of the surface topography at nanometer precision. For this reason, QPI has been demonstrated for the inspection of various samples, including nanopatterns with nanoscale defects [6, 7] and microelectromechanical systems [8-12], and for the real-time characterization of spatial light modulators [13] and photochemical etching processes [14]. However, despite these advantages, the interferometric nature of QPI methods makes them susceptible to environmental conditions, such as external vibrations and temperature fluctuations, limiting the stability of topographic measurements [15, 16].

Here, we demonstrate single-shot full-field topography measurements using spectrally multiplexed Kramers–Kronig holographic imaging. The intensity images at three different incident angles were simultaneously measured using a color imaging sensor, from which the phase images were retrieved based on Kramers–Kronig relations. Then, a height map of the reflective sample was reconstructed from the retrieved phase images using synthetic aperture holography. We imaged nanometer-scale vertical features (patterned silicon coated with gold) and cross-validated our results with AFM. The spatial and temporal sensitivities of our system were measured to be 1.80 nm and 0.36 nm, respectively.

A schematic of the experimental setup is shown in Figure 1(a). Three fiber-pigtailed laser diodes with center wavelengths of 658 nm, 520 nm, and 450 nm (Thorlabs Inc., LP660-SF20, LP520-SF15, and LP450-SF15) were used as light sources, which correspond to the red (R), green (G), and blue (B) channels, respectively. An aluminum plate was used to compactly align the three light sources (Fig. 1(a), left inlet). The illumination beams were then collimated and polarized by a series of lenses (L1–L3) and a linear polarizer. The combination of a polarizing beam splitter (PBS) and an achromatic quarter-wave plate separated the illumination and detection light paths in the reflection geometry. After the PBS, the collimated and linearly polarized illumination plane waves were projected onto the sample using a tube lens (L4) and an objective lens with a numerical aperture (NA) of 1.1 (Olympus Inc., LUMFL60XW). The sample was then imaged using the objective and tube lens and an additional 4$f$ telecentric imaging system (L5). Achromatic lenses (L1–L6) were employed in the setup to reduce optical dispersion. A motorized iris (Standa Ltd., 8MID30-1.5-N) was introduced at the Fourier plane between L5 and L6 to finely adjust the optical system NA, which was set to 0.99 (Fig. 1(a), the right inlet). Using a charge-coupled device with an RGB Bayer filter (XIMEA, MD120CU-SY), a color image was obtained in a single exposure and was decomposed into the three R, G, and B channel images (Fig. 1(b)). To correct the spectral cross-talk of the camera resulting from imperfect spectral filtering of the RGB Bayer filter, a color correction approach was utilized without interpolation [17].

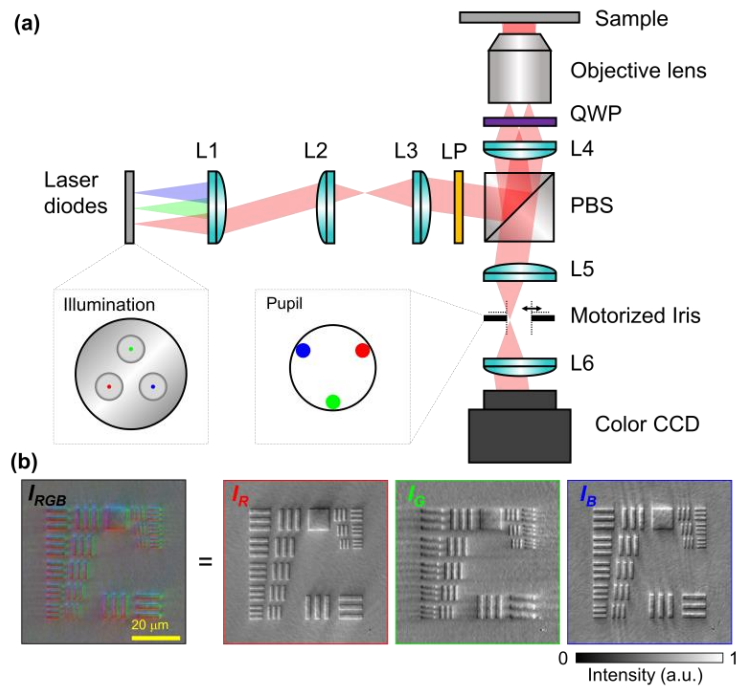

Fig. 1. Spectrally multiplexed Kramers–Kronig holographic imaging. (a) Experimental setup. L1– L6, achromatic lenses; L1, $f$ = 300 mm; L2, $f$ = 100 mm; L3–L6, $f$ = 200 mm; LP, linear polarizer; PBS, polarizing beam splitter; QWP, achromatic quarter-wave plate. (b) Spectral multiplexing. A color image is decomposed into three intensity images with respective R, G, and B channels.

The principle of the present method is based on [18], where the complex amplitude of the optical field can be obtained from a single intensity measurement using space-domain Kramers–Kronig relations. In order to use space-domain Kramers-Kronig relations, two conditions should be satisfied. First, the transverse wavevector of the incident plane wave $\mathbf{k}_{inc}$ must be the cut-off spatial angular frequency of the pupil function. Second, the intensity of the incident beam must be stronger than that of the light scattered from the sample. The first condition is experimentally realized by matching the angle of incidence with the effective NA of the objective lens. The second condition is satisfied when imaging a weakly scattering object. The resultant complex amplitude of the scattered field, $E$, is expressed in terms of the intensity:

$$E(\mathbf{r}) = \exp\left(\left\{\frac{\log[I(\mathbf{r})]}{2} - \frac{i}{\pi} P \int_{-\infty}^{\infty} \frac{\log[I(\mathbf{r}')]}{2(r'_\parallel - r_\parallel)} dr'_\parallel \right\} + i\mathbf{k}_{inc} \cdot \mathbf{r}\right) \quad (1)$$

where $\mathbf{r} = x\hat{\mathbf{x}} + y\hat{\mathbf{y}} = r_\parallel \hat{\mathbf{r}}_\parallel + r_\perp \hat{\mathbf{r}}_\perp$, $r = |\mathbf{r}|$ and $I$ is the measured intensity, $I = |E(\mathbf{r})|^2$, $P$ is the Cauchy principal value.

To verify the present method, we used a gold-coated KAIST logo pattern as a reflective sample with a height of 27 nm. From the measured intensity images using three different center wavelengths, the corresponding amplitude and phase images were retrieved using Eq. (1) (Fig. 2(a)). Then, the topographic image of the sample was reconstructed by synthesizing the retrieved optical fields in the Fourier space based on synthetic aperture holography [19] (Figs. 2(b)–2(c)). In this step, the retrieved fields were numerically refocused if necessary [20], and phase images of the respective R, G, and B channels were normalized to the mean wavelength $\lambda_c$ [21]. The synthesized phase image $\phi_{syn}(x, y)$ was then converted into a height map $h(x, y)$ as shown in Fig. 2(d), based on the following expression:

$$h(x, y) = \frac{\lambda_c}{4\pi n_m \cos\theta} \phi_{syn}(x, y) \quad (2)$$

where $n_m$ is the refractive index of the medium, and $\theta$ is the angle of incidence. In Eq. (2), the cosine factor accounts for the optical path differences induced by the reflection geometry, which is unity in the conventional epi-illumination scheme such as in [22].

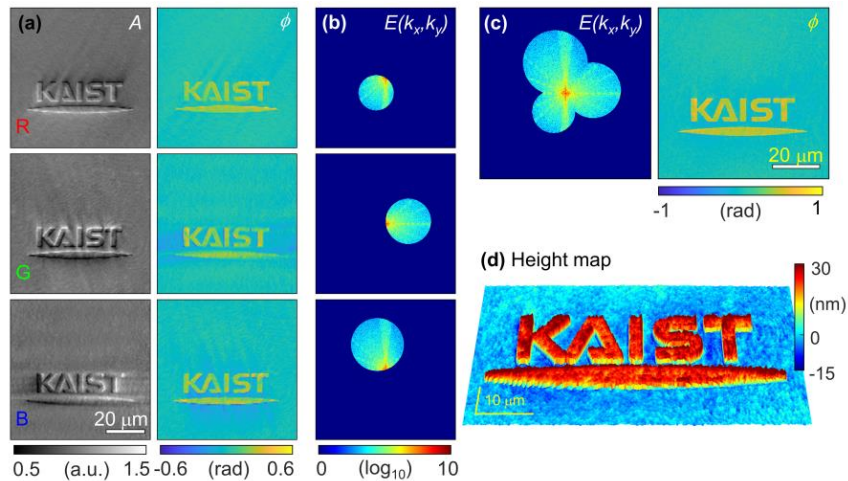

Fig. 2. Reconstructed the height map of a reflective logo pattern. (a) The retrieved amplitude $A$ and phase $\phi$ images of the pattern for respective R, G, and B channels. (b) Fourier transformed images of (a). (c) Synthetic results. By stitching the retrieved fields in the Fourier space, the single-shot high-resolution phase image of the sample is obtained. (d) The height map was obtained from the phase image in (c).

Although three light sources with different center wavelengths were used, correct heights are measured regardless of the optical dispersion of the sample because the measurement is conducted in reflection geometry. Additionally, all the refractive optical lenses used in the system were achromatic lenses.

Gold-coated 5-μm-width nanopillars, with heights ranging from 10 to 60 nm, were measured for validation using both the present method and AFM (Park Systems Inc., XE-100). Figures 3(a)–3(c) show the height maps of the nanopillars using our method, which agree well with the AFM measurements (Figs. 3(d)–3(f)). The line profiles of the marked lines in Fig. 3(a)-3(f) also confirmed that the results of Kramers–Kronig holographic imaging (Fig. 3(g), solid lines) were consistent with that of AFM (Fig. 3(g), dashed lines). In addition to the representative line profiles from the topography images, the two inspection methods were also compared using height histograms of the entire height maps. Figure 3(h) shows the height histograms of each nanopillar measured using the present method (colored distribution) and AFM (gray distribution), where the distance between the two peak values was considered as the nanopillar height. The peak values correspond to the representative heights of the background and nanopillar regions. In particular, the measured heights using the present method were 15.50 nm, 27.50 nm, and 57.50 nm, which were consistent with the measured heights using AFM: 14.25 nm, 27.00 nm, and 56.75 nm, respectively.

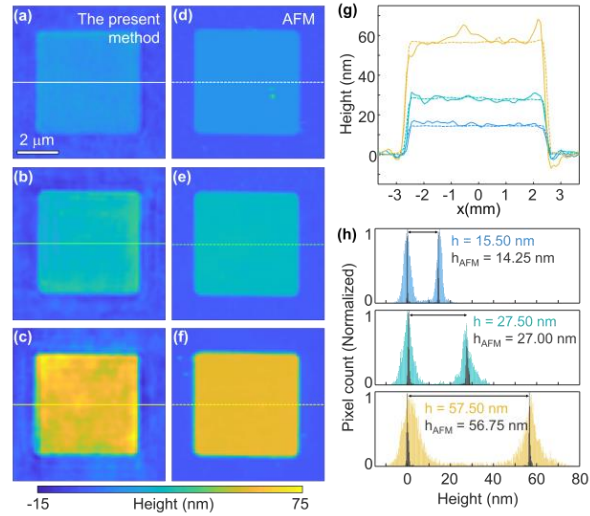

Fig. 3. Height map analyses of 5-μm width nanopillars of various heights. Height maps obtained using (a)–(c) the present method and (d)–(f) AFM. (g) Line profiles of the marked lines in (a)–(f). The line profiles confirm that the results (solid lines) are consistent with the AFM results (dashed lines). (h) Height histograms using the present method (colored distribution) and the AFM (gray distribution). For display purposes, each histogram is normalized by dividing the distribution by the maximum pixel count.

To demonstrate the full-field measurement capability of the present method, we measured 27-nm-height gold-coated samples with various patterns, including the USAF 1951 resolution target (Group 8–9), the KAIST logo pattern, and 5-μm-width nanopillars. The height maps of three samples were reconstructed from the single-shot full-field measurements, in which different patterns with the same height were clearly distinguished (Fig. 4). These results confirm that the present method is capable of characterizing the surface topography of both periodic and non-periodic structures with a large field of view.

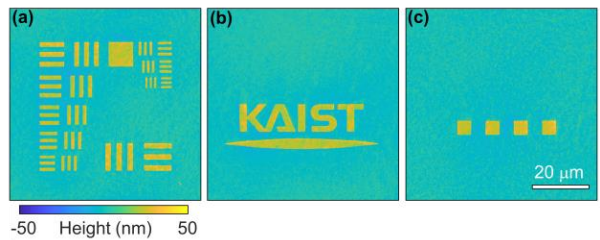

Fig. 4. Topographic images of 27-nm height reflective samples with various patterns: (a) the USAF resolution target; (b) the KAIST logo pattern; (c) 5-μm-width nanopillars.

Further, the system noise was characterized by measuring a region without a nanopillar (Fig. 5). The spatial noise was quantified as the standard deviation of the height map (Fig. 5(a)), while the temporal noise was measured as the spatial average of the temporal standard deviation for a series of 100 frames (Fig. 5(b)). The spatial and temporal noises were measured as 1.80 nm and 0.36 nm, respectively, which is comparable to that of previous 2D QPI methods [13, 15, 22, 23].

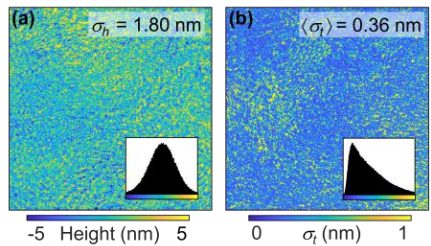

Fig. 5. Characterization of the system noise. (a) Measured spatial noise. (b) temporal noise.

In summary, we demonstrated single-shot full-field topography measurement using spectrally multiplexed Kramers–Kronig holographic imaging. We measured various nanometer-scale vertical features on gold-coated flat surfaces, and the results were compared with AFM measurements. The spatial and temporal noises were measured to be 1.80 and 0.36 nm, respectively.

Compared to existing holographic techniques, the single-shot and non-interferometric features of the present method allows for a simpler instrumentation and more stable measurements. The absence of a separate reference beam simplifies the instrumentation, and the use of temporally low-coherence light sources in single-path geometry ensures high spatiotemporal stability [24]. The laser diodes can also be replaced with light sources with even lower temporal coherence to further suppress the spatiotemporal noise of the system.

Accurate reconstruction of the surface topography using the principle of Kramers-Kronig relations is possible when the angles of incidence precisely match the optical system NA. It is also important that spatial coherence of the illumination must be ensured [18]. Because the present method utilizes the spectral multiplexing, the weak scattering assumption should be satisfied at the shortest wavelength of all the wavelengths used.

We envision that the present method could benefit various applications where rapid and precise surface topography is required, including real-time monitoring of the fabrication process of nanoscale structures and dynamic imaging of biomolecules and structures.

**Funding.** KAIST Up program; BK21+ program; Tomocube; National Research Foundation of Korea (2015R1A3A2066550); Institute of Information & communications Technology Planning & Evaluation (IITP) grant (2021-0-00745)

**Disclosures.** The authors declare that there are no conflicts of interest related to this article.

**Data availability.** Data underlying the results in this paper may be obtained from the authors upon request.

**Supplemental document**. There is no supplemental document.